\documentclass[12pt]{IEEEtran}
\usepackage{amsmath,graphicx,cite,bm,amssymb,amsthm,enumerate,epsfig,psfrag,cases,mathtools}
\renewcommand{\(}{\left(}
\renewcommand{\)}{\right)}
\renewcommand{\[}{\left[}
\renewcommand{\]}{\right]}
\renewcommand{\c}{\mathbf{c}}

\renewcommand{\b}{\mathbf{b}}

\newcommand{\f}{\mathbf{f}}

\renewcommand{\L}{\mathbf{L}}

\newcommand{\W}{\mathbf{W}}

\newcommand{\1}{\mathbf{1}}

\newcommand{\x}{\mathbf{x}}

\newcommand{\A}{\mathbf{A}}
\newcommand{\T}{\mathbf{T}}

\newcommand{\M}{\mathbf{M}}
\newcommand{\N}{\mathbb{N}}

\renewcommand{\u}{\mathbf{u}}
\renewcommand{\v}{\mathbf{v}}
\newcommand{\Q}{\mathbf{Q}}

\newcommand{\B}{\mathbf{B}}

\renewcommand{\b}{\mathbf{b}}

\newcommand{\Tr}[1]{{\rm{Tr}}\left(#1\right)}

\newcommand{\End}[1]{{\rm{End}}}

\renewcommand{\log}[1]{{\rm{log}}#1}

\newcommand{\var}[1]{{\rm{var}}\[#1\]}

\newtheorem{lemma}{Lemma}

\newtheorem{definition}{Definition}
\newtheorem{theorem}{Theorem}
\newtheorem{prop}{Proposition}

\newtheorem{rem}{Remark}

\usepackage[ruled,vlined,resetcount]{algorithm2e}
\usepackage{algorithmic,subfig}
\usepackage{fontenc}
\usepackage{inputenc}
\usepackage[square,sort,compress,comma,numbers]{natbib}
\usepackage{epstopdf,pst-node}
\usepackage{MnSymbol}
\usepackage{mathtools}
\usepackage{flushend}

\newcommand{\mypm}{\mathbin{\smash{%
\raisebox{0.35ex}{%
            $\underset{\raisebox{0.5ex}{$\smash -$}}{\smash+}$%
            }%
        }%
    }%
}

\DeclarePairedDelimiter\ceil{\lceil}{\rceil}
\DeclarePairedDelimiter\floor{\lfloor}{\rfloor}

\begin{document}
\title{Pseudo-Wigner Matrices}

\IEEEspecialpapernotice{\hfill\textit{Dedicated to the memory of Solomon W. Golomb}}

\author{Ilya Soloveychik, Yu Xiang and Vahid Tarokh, \\ John A. Paulson School of Engineering and Applied Sciences, \\ Harvard University
\thanks{This work was supported by the Fulbright Foundation and Army Research Office grant No. W911NF-15-1-0479.}
}
\maketitle

\begin{abstract}
We consider the problem of generating pseudo-random matrices based on the similarity of their spectra to Wigner's semicircular law. We introduce the notion of an $r$-independent pseudo-Wigner matrix ensemble and prove closeness of the spectra of its matrices to the semicircular density in the Kolmogorov distance. We give an explicit construction of a family of $N \times N$ pseudo-Wigner ensembles using dual BCH codes and show that the Kolmogorov complexity of the obtained matrices is of the order of $\log(N)$ bits for a fixed designed Kolmogorov distance precision. We compare our construction to the quasi-random graphs introduced by Chung, Graham and Wilson and demonstrate that the pseudo-Wigner matrices pass stronger randomness tests than the adjacency matrices of these graphs (lifted by the mapping $0 \rightarrow 1$ and $1 \rightarrow -1$) do. Finally, we provide numerical simulations verifying our theoretical results.
\end{abstract}


\section{Introduction}
Study of random matrices has been a very active area of research for the last few decades and has found enormous applications in various areas of modern mathematics, physics, engineering, biological modeling, and other fields \cite{akemann2011oxford}. In this article, we focus on square symmetric matrices with $\mypm 1$ entries, referred to as square symmetric {\it sign} matrices. For such matrices, Wigner \cite{wigner1955characteristic} demonstrated that if the elements of the upper triangle of an $N \times N$ square symmetric matrix are independent Rademacher ($\mypm 1$ with equal probabilities) random variables, then as $N$ grows, a properly scaled empirical spectral measure converges to the semicircular law (Wigner originally showed convergence in expectation, which was later improved to convergence in probability \cite{grenander2008probabilities} and to almost sure weak convergence \cite{arnold1971wigner,arnold1967asymptotic}). 

In many engineering applications, one needs to simulate random matrices. The most natural way to generate an instance of a random $N \times N$ sign matrix is to toss a fair coin $\frac{N(N+1)}{2}$ times, fill the upper triangular part of a matrix with the outcomes and reflect the upper triangular part into the lower. Unfortunately, for large $N$ such approach would require a powerful source of randomness due to the independence condition \cite{gentle2013random}. In addition, when the data is generated by a truly random source, atypical  \textit{non-random looking} outcomes have non-zero probability of showing up. Yet another issue is that any experiment involving tossing a coin would be impossible to reproduce. All these reasons stimulated researchers and engineers from different areas to seek for approaches of generating \textit{random-looking} data usually referred to as \textit{pseudo-random} sources or sequences of binary digits \cite{zepernick2013pseudo, golomb1967shift}. A wide spectrum of pseudo-random number generating algorithms have found applications in a large variety of fields including radar, navigation systems, digital signal processing, CDMA, error correction, cryptographic systems, Monte Carlo simulations, scrambling, coding theory, etc. \cite{zepernick2013pseudo}.

The term \textit{pseudo-random} is used to emphasize that the binary data at hand is indeed generated by an entirely deterministic causal process with low algorithmic complexity, but its statistical properties resemble some of the properties of data generated by tossing a fair coin. Remarkably, most efforts were focused on one dimensional pseudo-random sequences \cite{zepernick2013pseudo, golomb1967shift} due to their natural applications and to the relative simplicity of their analytical treatment. One of the most popular methods of generating pseudo-random sequences is due to Golomb \cite{golomb1967shift} and is based on linear-feedback shift registers capable of generating pseudo-random sequences of very low algorithmic complexity. The study of pseudo-random arrays and matrices was launched around the same time \cite{reed1962note, macwilliams1976pseudo, imai1977theory, sakata1981determining}. Among the known two dimensional pseudo-random constructions the most popular are the so-called perfect maps \cite{reed1962note, paterson1994perfect, etzion1988constructions}, and two dimensional cyclic codes \cite{imai1977theory, sakata1981determining}. However, none of these works considered spectral properties as the defining statistical features for their constructions.

There exist various approaches to quantifying the algorithmic power needed to generate an individual piece of binary data, also known as algorithmic complexity \cite{grunwald2004shannon, li2009introduction, downey2010algorithmic}. It can be intuitively thought of as a measure of amount of randomness stored in that piece of data. Below we stick to the concept of Kolmogorov complexity \cite{solomonoff1964formal1, kolmogorov1965three}. Let $D$ be a string of binary data of length $n$, then its Kolmogorov complexity is the length of the shortest binary Turing machine code that can produce $D$ and halt. If $D$ has no computable regularity it cannot be encoded by a program shorter than its original length $n$ (here and below we are mainly interested in the rate of growth of the Kolmogorov complexities of a sequence of constructions with their sizes, therefore, the complexities are usually given up to additive constants), meaning that its consecutive bits are unpredictable given the preceding ones and it may be considered as truly random \cite{knuth1998art, li2009introduction}. A string with a regular pattern, on the other hand, can be recovered by a program much shorter than the string itself, thus having a much smaller Kolmogorov complexity. By convention, a comparison of Kolmogorov complexities of various strings of the same length is usually done by conditioning on the length and thus assuming the length to be already known to the machine without specifying it as an input \cite{cover2012elements}. For example, the conditional Kolmogorov complexity of a Golomb sequence of length $n$ is at most $2\log_2 n$ \cite{golomb1967shift}, which is relatively small, since using a simple combinatorial argument one can show that at most $\frac{n}{2^n}$ fraction of the strings of length $n$ have conditional Kolmogorov complexity less than $\log_2 n$.

Specific pseudo-random sequence and array constructions usually start from a set of properties mimicking truly random data, and attempt to come up with deterministic ways of reproducing these properties. Following this approach, given a precision parameter $\varepsilon \geqslant \frac{1}{\log_2 N}$, we propose an explicit construction of scaled $N \times N$ symmetric sign matrices with high probability possessing spectra within $\varepsilon$-vicinity (in the Kolmogorov distance) of the semicircular law and having conditional Kolmogorov complexity proportional to $\frac{1}{\varepsilon}\log_2 N$.

The main contributions of this paper are as follows. First, we introduce the concept of an $r$-independent pseudo-Wigner ensemble and prove closeness of the spectra of its matrices to the semicircular law. Second, we give an explicit deterministic construction of such matrices which may replace random matrix generators in engineering applications. Third, using this construction we provide an upper bound on the amount of randomness needed to obtain Wigner's semicircular property, and show that it is surprisingly low. We also compare the proposed concept of pseudo-Wigner matrices with incidence matrices of the quasi-random graphs suggested by Chung, Graham and Wilson \cite{chung1989quasi} (lifted by the mapping $0 \rightarrow 1$, and $1 \rightarrow -1$) and show that our construction passes wider tests for randomness.

The outline of this paper is given next. In Section \ref{sec:notations}, we start with setting the notations and discussing a number of auxiliary results. We define $r$-independent pseudo-Wigner ensembles and analyze their spectral properties in Section \ref{sec:pwe}. In Section \ref{seq:exp_const} we provide an explicit construction of such matrices from dual BCH codes. We analyze the Kolmogorov complexity of the semicircular law in Section \ref{sec:kolm}. The relation of the pseudo-Wigner matrices to the quasi-random graphs is studied in Section \ref{sec:disc}. We support our theoretical findings by numerical experiments in Section \ref{sec:num} and provide our conclusions and final remarks in Section \ref{sec:conc}.

\section{Notation} 
\label{sec:notations}
For a vector $\x \in \mathbb{R}^N$, let $\|\x\|$ denote its Euclidean norm, and for a real symmetric matrix $\M \in \mathbb{R}^{N \times N}$, we denote its spectral norm by
\begin{equation}
\|\M\| := \max_{\x \in \mathbb{R}^N,\|\x\|=1}\x^T \M \x.
\end{equation}
For a real $x,\; \floor*{x}$ stands for the largest integer not exceeding $x$, and $\ceil*{x}$ stands for the smallest integer not less than $x$.

\noindent \textbf{Random variables.} For a real random variable $X$, we denote by $F_X(x)$ its cumulative distribution function (c.d.f. and c.d.f.-s in plural) and by $f_X(x)$ the corresponding probability density function (p.d.f.), The $p$-th central moment of $X$ is denoted by $M_p(X) := \mathbb{E}\[(X-\mathbb{E}X)^p\]$ when it exists. The second moment will also be written as $\var{X} := M_2(X)$. If a sequence of random variables $X_N$ convergences in distribution to a law $\mathcal{F}$, we write $X_N \xrightarrow[]{D} \mathcal{F}$.

We denote by $S_N$ the set of all symmetric $N \times N$ matrices with entries $\mypm \frac{1}{2\sqrt{N}}$. The Wigner ensemble $\mathcal{W}_N$ is defined as the set $S_N$ endowed with the uniform probability measure.

\noindent \textbf{Binary linear codes.} Let $\mathcal{C}$ be an $[n,k,d]$ binary linear code of length $n$, dimension $k$ and minimum Hamming distance $d$ over the field $GF(2)$. We say that two codewords $\u = \{u_i\},\; \v = \{v_i\} \in GF(2)^n$ are orthogonal if $\sum_i v_iu_i = 0$ in $GF(2)$. The dual code $\mathcal{C}^\perp$ of $\mathcal{C}$ is a linear code of length $n$ and dimension $k^\perp = n-k$, whose codewords are orthogonal to all the codewords of $\mathcal{C}$.

Let
\begin{equation}
\label{eq:zeta_def}
\begin{array}{llcl}
\hspace{-0.25 cm} \zeta_n : & GF(2)^n & \to & \{-1,1\}^n, \\
& \{u_i\}_{i=1}^n & \mapsto & \{(-1)^{u_i}\}_{i=1}^n,
\end{array}
\end{equation}
map binary $0/1$ sequences into sign sequences of the same length. Below whenever it does not lead to a confusion, we suppress the subscript and write $\zeta$ for simplicity.

\section{Pseudo-Wigner Ensembles} 
\label{sec:pwe}

For any symmetric matrix $\A_N \in S_N$, denote by $F_{\A_N}$ the c.d.f. associated with its real spectrum $\{\lambda_i\}_{i=1}^N$,
\begin{equation}
F_{\A_N}(x) = \frac{1}{N}\sum_{i=1}^n \theta(x-\lambda_i),
\end{equation}
where $\theta(x)$ is the unit step function at zero. The $l$-th moment of $\A_N$ is given by
\begin{equation}
\label{eq:mom_def_trace}
\beta_l(\A_N) = \int x^l dF_{\A_N} = \frac{1}{N}\Tr{\A_N^l}.
\end{equation}

Let $F_{sc}$ be the c.d.f. of the standard semicircular law
\begin{align}
\label{eq:sc_cdf_def}
&F_{sc}(x) \\ 
&\;\;= \begin{cases} \qquad\qquad\qquad 0, & x < -1, \\ \frac{1}{2} + \frac{1}{\pi}x\sqrt{1-x^2}+\frac{1}{\pi}\arcsin(x), & -1 \leqslant x \leqslant 1, \\ \qquad\qquad\qquad 1, & x > 1, \end{cases} \nonumber
\end{align}
with the corresponding p.d.f.
\begin{equation}
\label{eq:sc_pdf_def}
f_{sc}(x) = \begin{cases} \frac{2}{\pi} \sqrt{1-x^2},& -1 \leqslant x \leqslant 1, \\ \qquad 0, & \text{otherwise}. \end{cases}
\end{equation}
The moments of this distribution read as
\begin{equation}
\label{eq:dc_mom}
\beta_l = \int_{-\infty}^{+\infty} x^l dF_{sc} = \begin{cases} \quad\quad 0, & l \text{ odd}, \\ \frac{1}{2^l}\frac{l!}{\(\frac{l}{2}\)!\(\frac{l}{2}+1\)!}, & l \text{ even}.\end{cases}
\end{equation}

\subsection{The Semicircular Law and the Wigner Ensemble}
Recall that the Wigner ensemble $\mathcal{W}_N$ was defined to be the set $S_N$ of all $N \times N$ symmetric matrices with the entries $\mypm \frac{1}{2\sqrt{N}}$ endowed with the uniform probability measure.

\begin{lemma}[Main Theorem from \cite{sinai1998central}]
\label{lem:mom_conv_norm_or}
Suppose $\W_N \in \mathcal{W}_N$, then as $N \to +\infty$,
\begin{equation}
\mathbb{E}\[\beta_l(\W_N)\] = \begin{cases} \quad\;\; 0, & l \text{ odd}, \\ \beta_l + o(1), & l \text{ even},\end{cases}
\end{equation}
and the random variable $N(\beta_l(\W_N) - \mathbb{E}\[\beta_l(\W_N)\])$ converges in distribution to the normal law
\begin{equation}
\label{eq:norm_distr}
N(\beta_l(\W_N) - \mathbb{E}\[\beta_l(\W_N)\]) \xrightarrow[]{D} \mathcal{N}\(0,\frac{1}{\pi}\),
\end{equation}
in particular, 
\begin{equation}
M_p(N\beta_l(\W_N)) = \begin{cases} \qquad\; 0, & p \text{ odd}, \\ \frac{(p-1)!!}{\pi^{p/2}} + o(1), & p \text{ even}.\end{cases}
\end{equation}
\end{lemma}
This result in particular implies almost sure weak convergence of the empirical spectra of matrices from the Wigner ensemble to the semicircular law \cite{anderson2010introduction}.

\subsection{Pseudo-Wigner Ensemble}
Let us now introduce an ensemble of matrices matching the behavior of Wigner matrices up to a certain moment. Later we will show that if the number of matching moments grows logarithmically with the matrix size $N$, the empirical spectrum converges to the semicircular law with high probability.

\begin{definition}[$r$-independence of a sequence]
\label{def:ps_wig}
Let $\x = \{X_i\}_{i=1}^N$ be a sequence of centered sign-valued random variables. We say that $\x$ is $r$-independent if any $r$ of its elements $X_{i_1},\dots,X_{i_r}$ are statistically independent,
\begin{equation}
\mathbb{P}\[X_{i_1}=b_1,\dots,X_{i_r}=b_r\] = \prod_{l=1}^r \mathbb{P}\[X_{i_l}=b_l\],
\end{equation}
for any $i_1\neq\dots\neq i_r$ in the range $[1,N]$ and $b_i \in \{\mypm 1\}$.
\end{definition}

\begin{definition}[$r$-independent Pseudo-Wigner Ensemble of order $N$]
\label{def:ps_wig}
Let a subset $\mathcal{A}_N^r \subset S_N$ be endowed with the uniform measure. We say that it is an \textbf{$r$-independent pseudo-Wigner Ensemble of order $N$} if the elements of the upper triangular (including the main diagonal) parts of its matrices scaled by $2\sqrt{N}$ form an $r$-independent sequence w.r.t. (with respect to) the measure induced on them by $\mathcal{A}_N^r$.
\end{definition}

Below, whenever a probability over $\mathcal{A}_N^r$ is considered, it is always assumed to be w.r.t. to the uniform measure as in Definition \ref{def:ps_wig}. When the order $r$ is clear from the context, we suppress it and write $\mathcal{A}_N$. In addition, denote by
\begin{equation}
\beta_{l,N} = \mathbb{E}[\beta_l(\A_N)]
\end{equation}
the expected moments over ensemble $\mathcal{A}_N$. The next result justifies the title ``pseudo-Wigner'' in the above definition.
\begin{lemma}
\label{lem:mom_conv_norm}
Let $\gamma \in \N$ and $\A_N$ be chosen uniformly from $\mathcal{A}_N^{2\gamma r}$ with $r \geqslant l$. Then for the expected moments we have
\begin{equation}
\beta_{l,N} = \begin{cases} \quad\;\; 0, & l \text{ odd}, \\ \beta_l + o(1), & l \text{ even},\end{cases}
\end{equation}
as $N \to +\infty$. In addition, the first $p=1,\dots,2\gamma$ moments of the random variable $N(\beta_l(\A_N) - \beta_{l,N})$ converge to the moments of the normal law,
\begin{equation}
M_p(N\beta_l(\A_N)) = \begin{cases} \qquad\; 0, & p \text{ odd}, \\ \frac{(p-1)!!}{\pi^{p/2}} + o(1), & p \text{ even}.\end{cases}
\end{equation}
\end{lemma}

\begin{proof}
The proof follows that of the Main Theorem from \cite{sinai1998central} (Lemma \ref{lem:mom_conv_norm_or} above) and is based on counting paths of lengths up to $2\gamma l$ to calculate the corresponding empirical moments and their expectations. An essential ingredient of the proof consists of showing that the principal contribution to even moments is made by simple even paths (paths in which every edge is passed exactly twice), therefore, the resulting moments only depend on the variances of the matrix entries and not on their higher moments (universality). Due to the $2\gamma r$-independence with $r \geqslant l$, the calculation of the expected values of products of matrix variables on such paths (see formula (4.3) from \cite{sinai1998central}) will give exactly the same results. This completes the proof.
\end{proof}

Our next goal is to control the deviations of the spectra of matrices $\A_N \in \mathcal{A}_N$ from the semicircular law. For this purpose we use smoothing techniques based on finite polynomial expansions of the characteristic functions of the empirical and limiting distributions at hand.
\begin{lemma}[Lemma 7.4.2 from \cite{chung2001course}]
\label{lem:smoothing_ineq}
Suppose $F$ is a c.d.f. and $G: \mathbb{R} \to \mathbb{R}$ has bounded variation 
\begin{equation}
\int_{-\infty}^{+\infty}|G'(x)|dx < + \infty,
\end{equation}
bounded derivative
\begin{equation}
M = \sup_x |G'(x)| < + \infty,
\end{equation}
and satisfies
\begin{equation}
\lim_{x \to -\infty}G(x) = 0,\quad \lim_{x \to +\infty}G(x) = 1.
\end{equation}
Assume also that
\begin{equation}
\int_{-\infty}^{+\infty} |F(x)-G(x)|dx < +\infty,
\end{equation}
and denote by $\phi(t)$ and $\gamma(t)$ the Fourier transforms of $F(x)$ and $G(x)$ correspondingly. Then for any $T > 0$,
\begin{equation}
|F(x)-G(x)|\leqslant \frac{2}{\pi}\int_0^T \frac{|\phi(t)-\gamma(t)|}{t}dt + \frac{24M}{\pi T},
\end{equation}
uniformly over $x \in \mathbb{R}$.
\end{lemma}

\begin{theorem}
\label{th:main_res}
Let\footnote{Here we may consider a sequence of pseudo-Wigner ensembles.} $q < e$, $r \leqslant q\, \log_2 N$ and $\alpha \in (\frac{q}{e},1)$. Then  there exists $N_0$ such that for any $N \geqslant N_0$, with probability at least $1-\frac{r}{N^{2(1-\alpha)}}$ a matrix $\A_N$ chosen uniformly from $\mathcal{A}_N^{2r}$ satisfies
\begin{equation}
\label{eq:main_bound}
\left| F_{\A_N}(x) - F_{sc}(x)\right| \leqslant \frac{1}{r},\quad \forall x \in \mathbb{R}.
\end{equation}
\end{theorem}
\begin{proof}
The proof is based on the application of Chebyshev's inequality, and can be found in Appendix \ref{app:proof_main_res}.
\end{proof}

The same technique as in the proof of Theorem \ref{th:main_res} applied to higher moments yields
\begin{theorem}
\label{th:main_res_h}
Let $\gamma \in \N,\; q < e,\; r \leqslant q\, \log_2 N$ and $\alpha \in (\frac{q}{e},1)$. Then there exists $N_0$ such that for any $N \geqslant N_0$, with probability at least $1-\frac{3 r (2\gamma-1)!!}{(\sqrt{\pi} N^{1-\alpha})^{2\gamma}}$ a matrix $\A_N$ chosen uniformly from $\mathcal{A}_N^{2\gamma r}$ satisfies
\begin{equation}
\left| F_{\A_N}(x) - F_{sc}(x)\right| \leqslant \frac{1}{r},\quad \forall x \in \mathbb{R}.
\end{equation}
\end{theorem}
\begin{proof}
The proof is based on the high-moments version of Chebyshev's inequality and can be found in Appendix \ref{app:proof_main_res}.
\end{proof}

\section{An Explicit Construction of Pseudo-Wigner Matrices from BCH Codes}
\label{seq:exp_const}
We start this section by defining BCH codes, and briefly discuss the properties of their dual codes. Later we use the dual BCH codes to explicitly construct pseudo-Wigner ensembles.

\subsection{BCH Codes and Their Dual Codes}
\label{sec:code_descr}
We focus specifically on BCH codes for the following reasons:
\begin{itemize}
\item the construction of the BCH codes allows us to control their minimum distances in an easy manner, and
\item for relatively small designed minimum distances, the dimensions of the obtained BCH codes are close to maximal possible (see Section 1.10 from \cite{macwilliams1977theory} for details).
\end{itemize}
The importance of being able to control the minimum distance of a code is explained by Lemma \ref{lem:indep_source} below.

For $m \in \N$, a primitive narrow-sense binary BCH code $\mathcal{C}_m^\delta$ of length $n=2^m-1$ and designed minimum distance $\delta \geqslant 3$ is a cyclic code generated by the lowest degree binary polynomial having roots $\alpha,\;\alpha^2,\dots,\alpha^{\delta-1}$, where $\alpha$ is a primitive element of $GF(2^m)$ \cite{macwilliams1977theory}. 

\begin{lemma}[Theorem 9.1.1 and Corollary 9.3.8 from \cite{macwilliams1977theory}]
\label{th:bch_dim}
A primitive narrow-sense binary BCH code $\mathcal{C}_m^\delta$ of length $n=2^m-1$ and designed distance $\delta=2t+1$ with $1 \leqslant 2t-1 < 2^{[m/2]}+1$
\begin{itemize}
\item has minimum distance $d$ at least $\delta$, and
\item has dimension $n-mt$.
\end{itemize}
\end{lemma}

Under the same assumptions as in Lemma \ref{th:bch_dim}, the dual BCH code has dimension $k^\perp = mt$ \cite{macwilliams1977theory}, and is also a cyclic code.

\begin{lemma}[Lemma 3.2 from \cite{babadi2011spectral}]
\label{lem:indep_source}
If a code $\mathcal{C}$ has minimum distance $d$, then its dual code $\mathcal{C}^\perp$ is $(d-1)$-independent (see Definition \ref{def:ps_wig}) w.r.t. to the uniform measure over its codewords. 
\end{lemma}

For $N \in \N$, let $m \in \N$ satisfy
\begin{equation}
\label{eq:m_def}
2^{m-1}-1 < \frac{N(N+1)}{2} \leqslant 2^m-1.
\end{equation}
As before, denote $n = 2^m-1$. Fix $\delta$ small enough (Lemma \ref{th:bch_dim}) and construct a BCH code $\mathcal{C}_m^\delta$, whose parameters would be $[n,\;n-\frac{(\delta-1)m}{2},\;d]$ with $d \geqslant \delta$. For every codeword in the dual code $\c = \{c_i\}_{i=1}^n \in \(\mathcal{C}_m^\delta\)^\perp$, let $\b = \zeta(\c)$. Construct an $N \times N$ matrix $\overline{\B}_N$ by filling its upper triangular part (including the main diagonal) with the first $\frac{N(N+1)}{2}$ elements of the obtained sequence $\b$ in any specific order (e.g. fill the upper triangular part row by row) and then reflect it w.r.t. to the main diagonal. Finally, scale matrix $\overline{\B}_N$ by the factor of $\frac{1}{2\sqrt{N}}$ to normalize it 
\begin{equation} 
\B_N = \frac{1}{2\sqrt{N}}\overline{\B}_N.
\end{equation}
By Lemma \ref{th:bch_dim}, the dimension of the dual code is
\begin{equation}
\dim{\(\mathcal{C}_m^\delta\)^\perp} = \frac{(\delta-1)m}{2},
\end{equation} and due to Lemma \ref{lem:indep_source} this construction gives us a set $\mathcal{B}_N^{d-1}$ of matrices $\B_N$, which endowed with the uniform probability measure becomes a $d-1$-independent pseudo-Wigner ensemble of order $N$ with $d - 1 \geqslant \delta - 1$. Since $\mathcal{B}_N^{d-1}$ is also a $\mathcal{B}_N^{\delta-1}$ ensemble and $d$ may not be known, below we denote it by $\mathcal{B}_N^{\delta-1}$ to simplify notations. 

For example, for $\delta=2m+1$ we obtain the following
\begin{prop}
There exists $N_0$ such that for any $N \geqslant N_0$, if a matrix $\B_N$ is chosen uniformly from $\mathcal{B}_N^{\ceil*{2\log_2 N}}$, then with probability at least $1-\frac{2 \log_2 N}{N^{6/5}}$,
\begin{equation}
\left| F_{\B_N}(x) - F_{sc}(x)\right| \leqslant \frac{1}{\log_2 N},\quad \forall x \in \mathbb{R}.
\end{equation}
\end{prop}
\begin{proof}
This follows directly from Theorem \ref{th:main_res} by setting $q=1$ and $\alpha = \frac{2}{5}$.
\end{proof}

\section{The Kolmogorov Complexity of the Semicircular Law}
\label{sec:kolm}
The standard computer scientific approach to quantify the amount of randomness contained in a piece of data $D$ also known as its algorithmic compressibility is based on calculating the length of a minimal program creating that data on a universal Turing machine. The length of the obtained binary code is referred to as the Kolmogorov complexity of the object and we denote it by $\mathcal{KC}(D)$. A more fair comparison of Kolmogorov complexities of various objects of the same size is usually achieved by conditioning on that size $\mathcal{KC}(D|\text{size})$, or in other words by assuming that it is already known to the machine \cite{cover2012elements}.

The notion of the Kolmogorov complexity is naturally defined for specific instances of data. Since a particular matrix might not have the semicircular spectrum but can only be close to it, our goal is to generalize and extend the concept of Kolmogorov complexity to classes of objects sharing a specific property or, in other words, to sets of objects. As an example, for $\varepsilon > 0$ let us consider the following property
\begin{equation}
\mathcal{P}(N,\varepsilon) = \{\A_N \in S_N \mid \sup_x |F_{\A_N}(x)-F_{sc}(x)|\leqslant \varepsilon\},
\end{equation}
which is the set of symmetric $\frac{1}{2\sqrt{N}}$-scaled sign matrices of order $N$ having spectra at most $\varepsilon$ far from the semicircular law in the Kolmogorov metric. A naturally arising question can be formulated as: ``What is the smallest Kolmogorov complexity of a matrix from this set?'' This is the length of the shortest binary program needed to construct an object of a specific size possessing the necessary property. We suggest to take this quantity as the measure of randomness, or complexity, of the property. Motivated by this intuition, we propose the following formal definition.
\begin{definition}
\label{def:kc_prop}
The Kolmogorov complexity of a finite set (property) $\mathcal{P}$ is defined as
\begin{equation}
 \mathcal{KC}(\mathcal{P}) = \min_{D \in \mathcal{P}} \mathcal{KC}(D).
\end{equation}
\end{definition}
The conditional Kolmogorov complexity is defined analogously.

Next we investigate the Kolmogorov complexity of the semicircular property. 
Given a binary polynomial $f(x)$ of degree $m$, we write
\begin{equation}
\hat{f}(x) = x^mf(x^{-1})
\end{equation}
for its reciprocal.
 
\begin{prop}
\label{lem:kc_preudo_wig}
For $\varepsilon \geqslant \frac{1}{\log_2 N}$, the conditional Kolmogorov complexity of the property $\mathcal{P}(N,\varepsilon)$ is bounded by
\begin{equation}
\label{eq:kc_b_r}
\mathcal{KC}(\mathcal{P}(N,\varepsilon)|N) \leqslant \frac{2}{\varepsilon}\log_2 N + c,
\end{equation}
where $c$ does not depend on $N$ or $\varepsilon$.

A matrix sampled uniformly from $\mathcal{B}_N^{\ceil*{\frac{2}{\varepsilon}}}$ provides an explicit construction with probability at least $1-\frac{1}{\varepsilon N^{6/5}}$.
\end{prop}
\begin{proof}
Theorem \ref{th:main_res} guarantees that if the parameters of a pseudo-Wigner ensemble are chosen appropriately as functions of $\varepsilon$, at least one of the matrices from that ensemble must lie in the set $\mathcal{P}(N,\varepsilon)$. We will use this observation to obtain the desired bound on the Kolmogorov complexity of the property $\mathcal{P}(N,\varepsilon)$. 

Our goal is, thus, to build a dual BCH code with the designed minimum distance of the original code $\delta = \delta(\varepsilon)$, construct a pseudo-Wigner ensemble of $N \times N$ matrices based on it, and specify one matrix from it. Algorithm \ref{algo} contains \textit{pseudo-code} implementing the described algorithm. Note that the constructed matrix belongs to a $\delta-1$ independent pseudo-Wigner ensemble and Theorem \ref{th:main_res}, therefore, bounds the discrepancy of the empirical and limiting c.d.f.-s with high probability.

The upper bound on the Kolmogorov complexity is obtained by bounding the length of the algorithm's description. Note that the description of the Initialization step in Algorithm \ref{algo} requires at most 
\begin{equation}
\log_2 2^m +\log_2 2^\frac{(\delta-1)m}{2} + c_1 = m + \frac{(\delta-1)m}{2} + c_1
\end{equation}
bits to define polynomials $f(x)$ and $v(x)$ (represented by binary coefficient vectors $\f$ and $\v$) \cite{cover2012elements}. At this stage, the algorithm copies the values of $\f$ and $\v$ into the memory and the remaining code accesses them by their addresses, therefore, the description of steps $1-11$ have constant complexity not depending on $m$ or $\delta$. Overall, we obtain the following upper bound on the Kolmogorov complexity
\begin{equation}
\mathcal{L} \leqslant \frac{(\delta+1)m}{2} + c,
\end{equation}
where $c$ does not depend on $m$ or $\delta$. Theorem \ref{th:main_res} implies that the relation between the precision and the designed minimum distance is $\delta \sim \frac{2}{\varepsilon}$, which together with $(\ref{eq:m_def})$ yields (\ref{eq:kc_b_r}).

Now set $r = \ceil*{\frac{1}{\varepsilon}}$ and invoke Theorem \ref{th:main_res} with $\alpha = \frac{2}{5}$ to construct the necessary pseudo-Wigner ensemble and obtain the desired statement.
\end{proof}

\begin{algorithm}[t]
 \caption{Pseudo-Wigner Matrix}
 \begin{algorithmic}[1]
 \label{algo}
 \renewcommand{\algorithmicrequire}{\textbf{Input:}}
 \renewcommand{\algorithmicensure}{\textbf{Output:}}
 \REQUIRE $\f \in GF(2)^m,\v \in GF(2)^{\frac{(\delta-1)m}{2}}$
 \ENSURE  $\B_N$, s.t. $|F_{\B_N}(x)-F_{sc}(x)|\leqslant \frac{2}{\delta-1}$.
 \\ \textit{Initialization} : \text{read $\f,\; \v$ and $\delta$ into memory};
 \STATE construct $f(x) = \sum_j f_jx^j,\; v(x) = \sum_j v_jx^j$,
 \STATE build a splitting field $\mathcal{F}$ of $x^{2^m-1}-1$, which is also a splitting field for $f(x)$;
 \STATE let $\alpha \in \mathcal{F}$ be any roof of $f(x),\; f(\alpha)=0$;
 \FOR{$j = 2\colon \delta-2$}
 \STATE \text{find the min. polynomial $f_{j}(x)$ of $\alpha^j \in \mathcal{F}$};
 \IF{$f_{j}(x) \nmid f(x)$}
 \STATE $f(x) \leftarrow f(x) f_{j}(x)$;
 \ENDIF
 \ENDFOR
 \STATE $h(x) \leftarrow \dfrac{x^{2^m-1}-1}{\hat{f}(x)}$;
 \STATE take the codeword $\c$, whose polynomial representation is $v(x)h(x)$ and build $\B_N$ as described in Section \ref{sec:code_descr};
 \end{algorithmic} 
 \end{algorithm}

Proposition \ref{lem:kc_preudo_wig} demonstrates that the Kolmogorov complexity of property $\mathcal{P}(N,\varepsilon)$ for moderate values of $\varepsilon$ is proportional to $\frac{1}{\varepsilon}\log_2 N$ and is relatively small.

\section{Quasi-random Graphs}
\label{sec:disc}
In this section, we compare the proposed pseudo-Wigner matrices with the adjacency matrices of quasi-random Graphs from \cite{chung1989quasi}. Given a symmetric binary adjacency matrix $\T_N = \{t_{ij}\}$ of an undirected graph on $N$ vertices, we apply to it $\zeta$ transformation from (\ref{eq:zeta_def}) to get a sign matrix $\Q_N = \{q_{ij}\}$ (in (\ref{eq:zeta_def}) $\zeta$ was defined on sequences, however, a generalization to matrices is straightforward). The relation between $\T_N$ and $\Q_N$ can be written as
\begin{equation}
\label{eq:t_m_def}
\T_N = \frac{1}{2}\(\1\cdot\1^T + \Q_N\),
\end{equation}
where $\1 = [1,\dots,1]^T$ is a column vector of height $N$. As shown in \cite{chung1989quasi}, if a graph\footnote{To be precise, we should talk about sequences of graphs and their corresponding adjacency matrices. However, following \cite{chung1989quasi} to simplify the notations of this section and to better convey the intuition behind the calculations we prefer to talk about single instances of graphs and matrices.} satisfies condition $P_3$ given on page 347, it is a quasi-random graph. This condition can be formulated in terms of the adjacency matrix $\T_N$ as following. Let $N \to +\infty$, then if
\begin{enumerate}
\item the number of non-zero elements in $\T_N$ is $\sum_{ij}t_{ij} = \frac{N^2}{2} + o(N^2)$ (we have $2$ in the denominator instead of $4$ because we count all the edges twice due to the symmetry of the adjacency matrix),
\item $\lambda_1(\T_N) = \frac{N}{2}+o(N)$,
\item $\lambda_2(\T_N) = o(N)$,
\end{enumerate}
then the underlying graph is quasi-random. 

To demonstrate that our pseudo-random matrices may serve as a source of quasi-random graphs with high probability, let us prove the following
\begin{lemma}
\label{lem:spectr_bo}
For a matrix $\A_N \in \mathcal{A}_N^d$ and any vector $\x \in \mathbb{R}^N,\; \|\x\|=1$,
\begin{equation}
\mathbb{P}\[\x^T\A_N\x \geqslant 1\]\leqslant \frac{1}{N}.
\end{equation}
\end{lemma}
\begin{proof}
The proof can be found in Appendix \ref{app:aux}.
\end{proof}

Recall that the spectral norm of a symmetric real matrix is defined as
\begin{equation}
\|\Q_N\| = \max_{\x,\|\x\|=1}\x^T\Q_N\x,
\end{equation}
and note that we need to multiply matrix $\A_N$ by $2\sqrt{N}$ to get a sign matrix $\Q_N = 2\sqrt{N}\A_N$. Now as a corollary of Lemma \ref{lem:spectr_bo}, we get
\begin{equation}
\label{eq:qp_bound}
\mathbb{P}\[\|\Q_N\| \geqslant 2\sqrt{N}\] \leqslant \frac{1}{N},
\end{equation}
and therefore with high probability $\|\Q_N\| = o(N)$. 

\begin{lemma}[Weyl's Theorem, \cite{weyl1912asymptotische}]
Suppose the ordered eigenvalues of real symmetric $N \times N$ matrices $\L$ and $\L + \mathbf{N}$ are $\lambda_1 \geqslant \dots \geqslant \lambda_N$ and $\nu_1 \geqslant \dots \geqslant \nu_N$ correspondingly, then
\begin{equation}
\max_i\|\lambda_i - \nu_i\| \leqslant \|\mathbf{N}\|.
\end{equation}
\end{lemma}

Note that relation (\ref{eq:t_m_def}) may be viewed as a perturbation of matrix $\frac{1}{2}\1\cdot\1^T$, whose eigenvalues are $\lambda_1 = \frac{N}{2}$ and $\lambda_2=\dots=\lambda_{N-1} = 0$. Weyl's theorem together with (\ref{eq:qp_bound}) immediately imply that with high probability,
\begin{equation}
\lambda_1(\T_N) = \frac{N}{2}+o(N), \quad \lambda_2(\T_N) = o(N).
\end{equation}
In addition, set $\x = \frac{1}{\sqrt{N}}\1$ to get
\begin{equation}
\mathbb{P}\[\sum_{ij}q_{ij} \geqslant 2N^{3/2}\] \leqslant \frac{1}{N},
\end{equation}
which implies that $\sum_{ij}q_{ij} = o(N^2)$, or 
\begin{equation}
\sum_{ij}t_{ij} = \frac{N^2}{2} + o(N^2),
\end{equation}
as required. Therefore, we see that matrices from a pseudo-Wigner ensemble $\mathcal{A}_N^2$ with high probability exhibit properties of quasi-random graphs.

This simple example sheds more light on the hierarchy of properties of random graphs/matrices. We can conclude that if we have a sequence of pseudo-Wigner matrices of growing dimensions with fixed independence order $d(N) = d$, then we are more or less in the case of quasi-random graphs. Quasi-random graphs can be hardly considered random, as highly-structured Paley graphs example demonstrates \cite{chung1989quasi}. A higher level of complexity is the semicircular law, where we require $d(N)$ to grow with the size of the matrices. It can be in fact easily shown that the rate of growth of $d(N)$ only affects the speed of convergence, but not the limiting spectral law.

\section{Numerical Simulations}
\label{sec:num}
\begin{figure}[!h]
\centering
\includegraphics[width=3.7in]{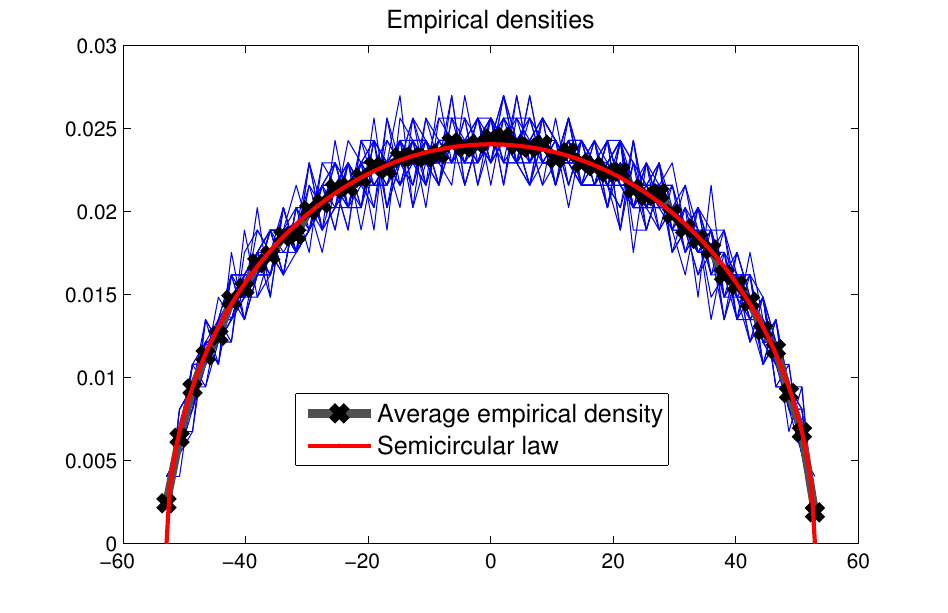}
\vspace{-0.5cm}
\caption{Empirical spectral densities of $25$ pseudo-Wigner matrices, $N = 700,\; m=18$.}
\label{fig}
\end{figure}
To demonstrate the power of the above construction, we chose $m=18,\; N=700$ and constructed an ensemble $\mathcal{B}_N^2$. In this case $d=3,\; t=1$ and the dual BCH code $\mathcal{C}$ is the so-called \emph{simplex} code, whose non-zero codewords are cyclic shifts of one Golomb sequence (see \cite{golomb1967shift} for details). A Golomb sequence may be easily generated by a linear-feedback shift register \cite{golomb1967shift} completely defined by picking a binary primitive polynomial of order $m$. Examples of binary primitive polynomials up to very high degrees can be found in tables, e.g. \cite{vzivkovic1994table}. In our simulation, we chose $f(x) = x^{18}+x^7+1$. Figure \ref{fig} shows the empirical spectra of $25$ matrices $\B_N \in \mathcal{B}_N^2$ together with their average spectrum in comparison with the semicircular law.

\section{Conclusions}
\label{sec:conc}
In this article, we consider the problem of defining and generating ensembles of pseudo-random matrices based on the similarity of their spectra to the semicircular density. We introduce $r$-independent pseudo-Wigner ensembles and prove their closeness to the semicircular law. We give an explicit construction of such ensembles using the dual BCH codes and compare them to the quasi-random graphs proposed by Chung, Graham and Wilson. We demonstrate that the Kolmogorov complexity of the proposed construction is proportional to $\frac{1}{\varepsilon}\log_2 N$, where $\varepsilon$ is a precision parameter, which is comparatively small. Finally, we justify the proposed construction numerically.

\appendices
\section{}
\label{app:proof_main_res}
\begin{proof}[Proof of Theorem \ref{th:main_res}]
The probability measures defined by the c.d.f.-s $F_{sc}(x)$ and $F_{\A_N}(x)$ are compact, therefore,
\begin{equation}
\int_{-\infty}^{+\infty} |F_{\A_N}(x)-F_{sc}(x)|dx < +\infty.
\end{equation}
The derivative of $F_{sc}$ is the semicircular p.d.f. $f_{sc}$ given in (\ref{eq:sc_pdf_def}), and is bounded by
\begin{equation}
\label{eq:m_bound}
M = \sup_x |f_{sc}(x)| = f_{sc}(0) = \frac{2}{\pi}.
\end{equation}
Recall also that the total variation of an almost surely differentiable c.d.f. is equal to one and apply Lemma \ref{lem:smoothing_ineq} to $F_{\A_N}(x)$ (in place of $F$) and $F_{sc}(x)$ (in place of $G$) to obtain
\begin{multline}
\label{eq:cdf_proofeq}
\left| F_{\A_N}(x) - F_{sc}(x)\right| \\ \leqslant \frac{2}{\pi}\int_0^T \frac{\left|\phi_{\A_N}(t)-\phi_{sc}(t)\right|}{t}dt + \frac{24M}{\pi T}.
\end{multline}
The $\rho-1$-th order Mclaurin polynomial expansion of the exponential function with a remainder gives the bound (see Section XV.4, pages 512-514 from \cite{feller1966introduction})
\begin{equation}
\label{eq:gam_b1}
\left| e^{ixt} - 1 - \sum_{l=1}^{\rho-1} \frac{(it)^l}{l!}x^l \right| \leqslant \frac{t^\rho}{\rho!}|x|^\rho,\;\; x, t \in \mathbb{R},\; t \geqslant 0.
\end{equation}
For even $\rho$, after taking expectations for fixed $t$ we get
\begin{equation}
\label{eq:gam_b2}
\left| \phi_{sc}(t) - 1 - \sum_{l=1}^{\rho-1} \beta_{l}\frac{(it)^l}{l!} \right| \leqslant \beta_{\rho}\frac{t^\rho}{\rho!}.
\end{equation}
Similarly,
\begin{equation}
\label{eq:phi_b}
\left| \phi_{\A_N}(t) - 1 - \sum_{l=1}^{\rho-1} \beta_{l}(\A_N)\frac{(it)^l}{l!} \right| \leqslant \beta_{l}(\A_N)\frac{t^\rho}{\rho!}.
\end{equation}
Set
\begin{equation}
\label{eq:def_rho}
\rho = \begin{cases} \quad r, & r \text{ even}, \\ 2\floor*{\frac{r-1}{2}}, & r \text{ odd}.\end{cases}
\end{equation}
Use (\ref{eq:gam_b2}) and (\ref{eq:phi_b}) to obtain the following bound on the integral summand of the right-hand side of (\ref{eq:cdf_proofeq}),
\begin{align}
\int_0^T & \left|\frac{\phi_{\A_N}(t)-\phi_{sc}(t)}{t}\right|dt \nonumber \\
& \leqslant \int_0^T \frac{1}{t}\left|\sum_{l=1}^{\rho-1} \beta_{l}(\A_N)\frac{(it)^l}{l!} - \sum_{l=1}^{\rho-1} \beta_{l}\frac{(it)^l}{l!}\right|dt  \nonumber \\ 
& \hspace{2 cm} + \int_0^T \[\beta_{\rho}(\A_N)+\beta_{\rho}\]\frac{t^{\rho-1}}{\rho!}dt \nonumber \\
& \leqslant
\int_0^T \frac{1}{t}\left|\sum_{l=1}^{\rho-1} \[\beta_{l}(\A_N) - \beta_l\] \frac{(it)^l}{l!}\right|dt  \nonumber \\ 
& \hspace{1 cm} + \[\beta_{\rho}(\A_N)+\beta_{\rho}\]\frac{T^\rho}{\rho\cdot \rho!} = S_1 + S_2. 
\label{eq:phi_ineq}
\end{align}
Bound $S_1$ as
\begin{multline}
\label{eq:s1_expr}
S_1 = \int_0^T \frac{1}{t}\left|\sum_{l=1}^{\rho-1} \[\beta_{l}(\A_N) - \beta_l\] \frac{(it)^l}{l!}\right|dt \\ \leqslant \max_{1\leqslant l \leqslant \rho-1} |\beta_{l}(\A_N) - \beta_l| \int_0^T \sum_{l=1}^{\rho-1}\frac{t^{l-1}}{l!}dt.
\end{multline}
From the triangle inequality,
\begin{equation}
\label{eq:proof_triang}
|\beta_{l}(\A_N) - \beta_l| \leqslant |\beta_{l,N} - \beta_l| + |\beta_{l}(\A_N) - \beta_{l,N}|.
\end{equation}
For the first summand we have
\begin{equation}
|\beta_{l,N} - \beta_l| \leqslant \frac{2\rho}{N},
\end{equation}
which follows directly from the path counting, see \cite{bai2010spectral} for details. 

By Lemma \ref{lem:mom_conv_norm}, the variance of the second summand of (\ref{eq:proof_triang}) satisfies
\begin{equation}
\var{\beta_{l}(\A_N) - \beta_{l,N}} = \frac{1}{\pi N^2}+o\(\frac{1}{N^2}\),\;\; N \to + \infty.
\end{equation}
Therefore, there exists $N_0 \in \N$ such that
\begin{equation}
\var{\beta_{l}(\A_N)- \beta_{l,N}} \leqslant \frac{2}{\pi N^2},\;\; \forall N \geqslant N_0.
\end{equation}
Now the Chebyshev's bound gives starting from $N_0$,
\begin{equation}
\label{eq:appl_cheb}
\mathbb{P}\[|\beta_{l}(\A_N) - \beta_{l,N}| \geqslant \delta\] \leqslant \frac{2}{\pi N^2 \delta^2}.
\end{equation}
Apply the union bound to the maximum in (\ref{eq:s1_expr}) to get
\begin{multline}
\mathbb{P}\[\max_{1\leqslant l \leqslant \rho-1} |\beta_{l}(\A_N) - \beta_l| \geqslant \delta\] \\ \leqslant \sum_{l=1}^{\rho-1} \mathbb{P}\[|\beta_{l}(\A_N) - \beta_l| \geqslant \delta\] \leqslant \frac{2 (\rho-1)}{\pi N^2 \delta^2}.
\end{multline}
Note that for any $a, b \in \N$,
\begin{equation}
\sum_{l=a}^b \frac{t^l}{l\cdot l!} \leqslant \sum_{l=0}^{+\infty} \frac{t^l}{l!} = e^t.
\end{equation}
We conclude that
\begin{equation}
\int_0^T \sum_{l=1}^{\rho-1}\frac{t^{l-1}}{l!}dt \leqslant \sum_{l=1}^{\rho-1} \frac{T^l}{l\cdot l!} \leqslant e^T.
\end{equation}
Overall, for $S_1$ we have
\begin{equation}
\mathbb{P}\[S_1 \geqslant \(\delta+\frac{2\rho}{N}\)e^T\] \leqslant \frac{2 (\rho-1)}{\pi N^2 \delta^2}.
\end{equation}
Choose
\begin{equation}
\delta = \frac{1}{N^{\alpha}},
\end{equation}
and let
\begin{equation}
\label{eq:t_choice}
T = \frac{\rho^{1+1/\rho}}{e}.
\end{equation}
Recall the assumptions: $\alpha < 1$ and
\begin{equation}
\label{eq:r_cond_b}
\rho \leqslant q\,\log_2 N,
\end{equation}
to get for $N \geqslant N_0$ large enough,
\begin{align}
\(\delta+\frac{2\rho}{N}\)e^T & \leqslant \(\frac{1}{N^{\alpha}}+\frac{2q\, \log_2 N}{N}\)e^{\rho/e}e^{\sqrt[\leftroot{-2}\uproot{2}\rho]{\rho}} \nonumber\\
& \leqslant \frac{1}{N^{\alpha}}N^{\frac{q}{e}}e^{\sqrt[\leftroot{-2}\uproot{2}\rho]{\rho}},
\end{align}
and therefore
\begin{equation}
\label{eq:ps1_be}
\mathbb{P}\[S_1 \geqslant \frac{e^{\sqrt[\leftroot{-2}\uproot{2}\rho]{\rho}}}{N^{\alpha-\frac{q}{e}}}\] \leqslant \frac{2 (\rho-1)}{\pi N^{2(1-\alpha)}}.
\end{equation}
Using the same Chebyshev's bound (\ref{eq:appl_cheb}) and the triangle inequality, for the second summand on the right-hand side of (\ref{eq:phi_ineq}) we obtain
\begin{equation}
\label{eq:ps2_est}
\mathbb{P}\[S_2 \geqslant \(2\beta_{\rho}+\eta+\frac{2\rho}{N}\) \frac{T^r}{\pi \rho\cdot \rho!}\] \leqslant \frac{2}{\pi N^2 \eta^2}.
\end{equation}
Use Stirling's approximation 
\begin{equation}
\label{eq:stirl}
\sqrt{2\pi}\rho^{\rho+\frac{1}{2}}e^{-\rho} \leqslant \rho! \leqslant e\rho^{\rho+\frac{1}{2}}e^{-\rho},
\end{equation}
to get from (\ref{eq:dc_mom}) the following bound
\begin{equation}
\beta_\rho = \frac{1}{2^\rho}\frac{\rho!}{\(\frac{\rho}{2}\)!\(\frac{\rho}{2}+1\)!} \leqslant \frac{2}{\rho^{3/2}}.
\end{equation}
Plug this result into (\ref{eq:ps2_est}), recall (\ref{eq:r_cond_b}), and set $\eta = \frac{1}{\rho^{3/2}}$ to get
\begin{equation}
\mathbb{P}\[S_2 \geqslant \frac{1}{\rho^{5/2}} \frac{T^\rho}{\cdot \rho!}\] \leqslant \frac{2\rho^3}{\pi N^2}.
\end{equation}
Now use (\ref{eq:t_choice}) to obtain
\begin{equation}
\frac{1}{\rho^{5/2}} \frac{T^\rho}{\cdot \rho!} \leqslant \frac{1}{\rho^2},
\end{equation}
and thus
\begin{equation}
\label{eq:ps2_be}
\mathbb{P}\[S_2 \geqslant \frac{1}{\rho^2}\] \leqslant \frac{2\rho^3}{\pi N^2}.
\end{equation}
Finally, using the inequality (\ref{eq:r_cond_b}) again, we get from (\ref{eq:ps1_be}) and (\ref{eq:ps2_be}) that for $N \geqslant N_0$ large enough,
\begin{multline}
\mathbb{P}\[S_1 + S_2 \geqslant \frac{2}{\rho^2}\] \leqslant \frac{2\rho^3}{\pi N^2} + \frac{2\rho}{\pi N^{2(1-\alpha)}} \\ \leqslant \frac{3\rho}{\pi N^{2(1-\alpha)}} \leqslant \frac{\rho}{N^{2(1-\alpha)}},
\end{multline}
where we have used the union bound and noted that in order for the sum $S_1+S_2$ to be greater than $\frac{2}{\rho^2}$, at least one of the summands must necessarily be greater than $\frac{1}{\rho^2}$. Altogether, with probability at least $1-\frac{\rho}{N^{2(1-\alpha)}}$,
\begin{equation}
\int_0^T \frac{1}{t}\left|\sum_{l=1}^{\rho-1} \[\beta_{l}(\A_N) - \beta_l\] \frac{(it)^l}{l!}\right|dt \leqslant \frac{2}{\rho^2}.
\end{equation}
Plug this bound and (\ref{eq:t_choice}) into (\ref{eq:cdf_proofeq}) to conclude that with probability at least $1-\frac{\rho}{N^{2(1-\alpha)}}$,
\begin{equation}
\label{eq:cdf_proofeq1}
\left| F_{\A_N}(x) - F_{sc}(x)\right| \leqslant \frac{2}{\rho^2} + \frac{24eM}{\rho^{1+1/\rho}}.
\end{equation}
According to (\ref{eq:m_bound}),
\begin{equation}
M = \frac{2}{\pi},
\end{equation}
therefore, for $N \geqslant N_0$ large enough,
\begin{equation}
\label{eq:cdf_proofeq1}
\left| F_{\A_N}(x) - F_{sc}(x)\right| \leqslant \frac{1}{\rho}.
\end{equation}
To conclude the proof we apply (\ref{eq:def_rho}).
\end{proof}

\begin{rem}
Note that unlike \cite{babadi2011spectral}, we cannot apply Lemma XVI.3.2 from \cite{feller1966introduction} to prove Theorem \ref{th:main_res}, since in our case the empirical spectral measures $F_{\A_N}(x)$ are in general not centered. Due to this distinction, we use another smoothing inequality given by Lemma \ref{lem:smoothing_ineq}.
\end{rem}

\begin{proof}[Proof of Theorem \ref{th:main_res_h}]
The proof goes along the lines of the proof of Theorem \ref{th:main_res} up to equation (\ref{eq:appl_cheb}), where we use a stronger version of Chebyshev's inequality for higher moments, namely,
\begin{equation}
\label{eq:cheb_gen}
\mathbb{P}\[|X - \mathbb{E}X|\geqslant \delta\] \leqslant \frac{M_p(X)}{\delta^p},
\end{equation}
where $X$ is a random variable with a finite $p$-th absolute moment 
\begin{equation}
\overline{M}_p(X) = \mathbb{E}\[|X-\mathbb{E}X|^p\].
\end{equation}
Since according to Lemma \ref{lem:mom_conv_norm}
\begin{equation}
N(\beta_l(\A_N) - \beta_{l,N}) \overset{D}{\to} \mathcal{N}\(0,\frac{1}{\pi}\),
\end{equation}
for $l \leqslant \rho$ with $\rho$ as in (\ref{eq:def_rho}), the absolute moments read as
\begin{multline}
\mathbb{E}\[|\beta_l(\A_N) -  \beta_{l,N}|^p\] = \frac{(p-1)!!}{(\sqrt{\pi} N)^p}\cdot \begin{cases} \sqrt{\frac{2}{\pi}}, & p \text{ even}, \\ \;\;\; 1, & p \text{ odd}\end{cases} \\ + o\(\frac{1}{N^p}\),\;\; n \to +\infty.
\end{multline}
Therefore, there exists $N_0$ such that
\begin{equation}
M_p(\beta_{l}(\A_N)) \leqslant \frac{2(p-1)!!}{(\sqrt{\pi} N)^p},\;\; \forall N \geqslant N_0.
\end{equation}
Now Chebyshev's bound (\ref{eq:cheb_gen}) implies that starting from $N_0$,
\begin{equation}
\label{eq:appl_cheb_new}
\mathbb{P}\[|\beta_{l}(\A_N) - \beta_l| \geqslant \delta\] \leqslant \frac{2(p-1)!!}{(\sqrt{\pi} N\delta)^p}.
\end{equation}
In our case we know that moments up to order $2\gamma$ must coincide with those of the Wigner ensemble, therefore, we get
\begin{multline}
\mathbb{P}\[\max_{1\leqslant l \leqslant \rho-1} |\beta_{l}(\A_N) - \beta_l| \geqslant \delta\] \leqslant \sum_{l=1}^{\rho-1} \mathbb{P}\[|\beta_{l}(\A_N) - \beta_l| \geqslant \delta\] \\
\leqslant \frac{2(\rho-1)(2\gamma-1)!!}{(\sqrt{\pi} N\delta)^{2\gamma}}.
\end{multline}
Following the proof of Theorem \ref{th:main_res} and using the bound in (\ref{eq:appl_cheb_new}) instead of (\ref{eq:appl_cheb}) we get the desired result.
\end{proof}

\section{}
\label{app:aux}
\begin{proof}[Proof of Lemma \ref{lem:spectr_bo}]
Introduce a random variable,
\begin{equation}
\xi = \x^T\A_N\x,
\end{equation}
then $\mathbb{E}[\xi] = 0$ and its variance reads as
\begin{equation*}
\var{\xi} = \mathbb{E}[(\x^T\A_N\x)^2] = \mathbb{E}\[\sum_{ijkl}x_ix_jx_kx_l a_{ij}a_{kl}\],
\end{equation*}
where $\A_N = \{a_{ij}\}_{i,j=1}^N$. Since the elements from the upper triangular part of $\A_N$ are pairwise independent,  the only contribution to this expectation is made by summands with $i=k,\; j=l$ and $i=l,\;j=k$. Recall also that $\mathbb{E}[a_{ij}^2]=\frac{1}{4N}$, to get
\begin{multline}
\var{\xi} =  2\mathbb{E}\[\sum_{ij}x_i^2x_j^2 a_{ij}^2\] =\frac{2}{4N}\sum_{ij}x_i^2x_j^2 \\
= \frac{1}{2N}\(\sum_{i}x_i^2\)\(\sum_{j}x_j^2\) = \frac{1}{2N}.
\end{multline}
Finally, Chebyshev's bound yields
\begin{equation}
\mathbb{P}\[\x^T\A_N\x \geqslant \delta\] \leqslant \frac{1}{2N\delta^2},
\end{equation}
and by setting $\delta=1$ we obtain the desired statement.
\end{proof}

\bibliographystyle{IEEEtran}
\bibliography{ilya_bib}
\end{document}